\documentclass[12pt]{article}
\pdfoutput=1
\usepackage{amssymb,amsmath, graphicx}
\usepackage{latexsym}
\usepackage{mathrsfs}
\usepackage{hyperref} 
\usepackage{verbatim}
\usepackage{bm}
\usepackage{cite}

\def\bea{\begin{eqnarray}}
\def\eea{\end{eqnarray}}
\def\be{\begin{equation}}
\def\ee{\end{equation}}
\def\ba{\begin{array}}
\def\ea{\end{array}}

\def\nn{\nonumber}

\newcommand{\calF}{{\cal F}}

\newcommand{\dph}{{\dot{\phi}}}
\newcommand{\dth}{{\dot{\theta}}}
\newcommand{\dvp}{{\dot{\varphi}}}

\begin{document}

\setlength\arraycolsep{2pt}

\renewcommand{\theequation}{\arabic{section}.\arabic{equation}}
\setcounter{page}{1}

\begin{titlepage}

\begin{center}

\vskip 1.0 cm

{\LARGE  \bf  Cosmic inflation in a landscape of heavy-fields}

\vskip 1.0cm

{\large
Sebasti\'an C\'espedes and  Gonzalo A. Palma
}

\vskip 0.5cm

{\it
Physics Department, FCFM, Universidad de Chile,\\ \mbox{Blanco Encalada 2008, Santiago, Chile}
}

\vskip 1.5cm

\end{center}

\begin{abstract}

Heavy isocurvature fields may have a strong influence on the low energy dynamics of curvature perturbations during inflation, as long as the inflationary trajectory becomes non-geodesic in the multi-field target space (the landscape). If fields orthogonal to the inflationary trajectory are sufficiently heavy, one expects a reliable effective field theory describing the low energy dynamics of curvature perturbations, with self-interactions determined by the shape of the inflationary trajectory. Previous work analyzing the role of heavy-fields during inflation have mostly focused in the effects on curvature perturbations due to a single heavy-field. In this article we extend the results of these works by studying models of inflation in which curvature perturbations interact with two heavy-fields. We show that the second heavy-field (orthogonal to both tangent and normal directions of the inflationary trajectory) may significantly affect the evolution of curvature modes. We compute the effective field theory for the low energy curvature perturbations obtained by integrating out the two heavy-fields and show that the presence of the second heavy-field implies the existence of additional self-interactions not accounted for in the single heavy-field case. We conclude that future observations will be able to constrain the number of heavy fields interacting with curvature perturbations.

\end{abstract}

\end{titlepage}

\newpage

\section{Introduction} \label{intro}
\setcounter{equation}{0}


Undoubtedly, canonical models of single-field slow-roll inflation\footnote{By canonical models of single-field slow-roll inflation we mean models derived from an action of the form $S = S_{\rm EH} - \int \left[ \frac{1}{2} (\partial \phi)^2 + V(\phi) \right]$ where $S_{\rm EH}$ is the usual Einstein-Hilbert term.} give us the simplest resolution to the horizon and flatness problems encountered in hot big-bang cosmology~\cite{Guth:1980zm, Linde:1981mu, Albrecht:1982wi}, and offer us an elegant explanation to the origin of primordial curvature perturbations, characterized by a nearly scale invariant power spectrum~\cite{ Mukhanov:1981xt}. Although such predictions are fully compatible with current observations~\cite{Komatsu:2010fb, Hlozek:2011pc, Bennett:2012fp}, there is still plenty of room for a change in paradigm in the advent of future experiments, such as large scale structure surveys~\cite{Abell:2009aa, Schlegel:2011wb, Amendola:2012ys} and 21cm cosmology~\cite{21cm}. A possible observation of scale dependence in the primordial spectra ({\it i.e.} in the form of features and/or running)~\cite{Starobinsky:1992ts, Polarski:1992dq, Chung:1999ve, Adams:2001vc, Martin:2004yi, Gong:2005jr, Shafieloo:2006hs, Ashoorioon:2006wc, Hamann:2007pa,  Ashoorioon:2008qr, Romano:2008rr, Tye:2008ef, Hamann:2009bz, Tye:2009ff, Barnaby:2010ke, Achucarro:2010da, Chen:2011zf, Park:2012rh, Cespedes:2012hu, Chluba:2012we, Battefeld:2013xka, Jackson:2013mka} and/or large non-Gaussianity~\cite{Linde:1996gt, Bartolo:2001cw, Bernardeau:2002jy, Maldacena:2002vr, Lyth:2002my, Alishahiha:2004eh, Bartolo:2004if, Chen:2006nt, Chen:2009we} would force us to leave this simple picture behind, and move on to consider models of inflation where the evolution of curvature perturbations was influenced by nontrivial self-couplings and/or interactions with additional degrees of freedom. 

Elucidating how future observations will guide our understanding of inflationary cosmology beyond the standard single-field paradigm has been the main focus of much effort during recent years~\cite{Review}. One particularly powerful and compelling framework to analyze inflation in a model independent way is the recently proposed effective field theory approach~\cite{Cheung:2007st} (see also \cite{Burgess:2003zw, Weinberg:2008hq}). In this scheme, the broken time translation invariance of the inflationary background is parametrized by introducing a Goldstone boson field $\pi (x,t)$, defined as the perturbation along the broken time translation symmetry.  At the same time, curvature perturbations are intimately related to the Goldstone boson, whose action appears highly constrained by the symmetries of the original ultraviolet (UV)-complete action. In particular, the unknown UV-physics is parametrized by self-interactions of the Goldstone boson that non-linearly relate field operators at different orders in perturbation theory. This framework has offered a powerful approach to analyze the large variety of infrared observables potentially predicted by inflation, including the prediction of non-trivial signals in the primordial power spectrum and bispectrum~\cite{Cheung:2007sv, Senatore:2009gt, Senatore:2009cf, Senatore:2010jy, Senatore:2010wk, Creminelli:2010qf, Baumann:2011dt, Baumann:2011su, Baumann:2011nk, Baumann:2011nm, LopezNacir:2011kk, Baumann:2011ws,Noumi:2012vr, Achucarro:2012sm, Senatore:2012wy, Gwyn:2012mw, Behbahani:2012be,  Achucarro:2012fd}. At short wavelengths, for instance, one finds that the Goldstone boson action is given by~\cite{Cheung:2007st}  
\be
S = - M_{\rm Pl}^2 \int d^4 x \dot H \left[ \frac{1}{c_s^2} \left( \dot \pi^2 - c_s^2 \frac{(\partial_i \pi)^2}{a^2} \right) - \frac{1-c_s^2}{c_s^2} \left( \frac{(\partial_i \pi)^2}{a^2} + \frac{A}{c_s^2} \dot \pi^2 \right) \dot \pi + \cdots \right] , \label{eft-basic}
\ee
where $c_s$ is the speed of sound at which Goldstone boson quanta propagate, and $A$ is a quantity that parametrizes different models of inflation (for instance, DBI inflation~\cite{Alishahiha:2004eh} corresponds to the particular case $A=-1$). Current available data~\cite{Bennett:2012fp} mildly constrain $c_s$ and $A$, suggesting that future observations might rule out a large variety of models of inflations.

Arguably, the simplest class of theories incorporating a departure from canonical single-field slow-roll inflation is offered by models in which adiabatic modes (or equivalently, Goldstone boson modes) interact with heavy scalar fields, with masses much larger than the expansion rate during inflation~\cite{Tolley:2009fg, Achucarro:2010jv, Achucarro:2010da, Cespedes:2012hu, Gao:2012uq, Pi:2012gf, Achucarro:2012yr, Saito:2012pd}. Crucially, such models continue to be of the single field type~\cite{Achucarro:2012yr}, but come dressed with properties that differ significantly from those encountered in standard single-field models. Indeed, near horizon crossing the Goldstone boson modes do not carry enough energy to excite their high-energy counterparts implied by the heavy-fields, meaning that curvature perturbations are generated by a single low energy degree of freedom. Nevertheless, the presence of heavy-fields can induce self-couplings for adiabatic perturbations that may have a sizable impact on their evolution (for example, by modifying the dispersion relation of the Goldstone boson mode). This has been understood gradually in a series of recent articles~\cite{Achucarro:2012sm, Achucarro:2012yr, Gwyn:2012mw}, and for the particular case of models with a Goldstone boson mode interacting with a single heavy-field\footnote{That is, in the particular case where the original theory consists of a two-scalar field model with a potential such that there is only one flat direction, followed by the inflationary trajectory.}, our current understanding may be summarized as follows:
\begin{itemize}
\item There exists a background inflationary trajectory which traverses the multi-field landscape determined by the scalar field potential of the theory. In general, this trajectory is expected to be non-geodesic, meaning that the flat directions of the scalar potential do not necessarily align with the family of geodesic paths defined by the scalar manifold of the theory's target space. It is possible to think of such non-geodesic trajectories as turning trajectories, characterized by an angular velocity~$\dot \theta$ (the rate of turn of the trajectory).

\item To study the perturbations of the system, it is useful to define perturbations along the trajectory and perpendicular to it. The first class defines the Goldstone boson field $\pi(t,x)$ and the second one corresponds to a heavy scalar field with an effective mass $M_{\rm eff}$ given by $M_{\rm eff}^2 = m^2 - \dot \theta^2$, where $m$ is the standard value of the mass computed from the potential alone. The angular velocity $\dot \theta$ is found to have an important role on the dynamics of these two perturbations, as it implies nontrivial interactions between the Goldstone boson and the heavy-field.

\item Because of these interactions, both the Goldstone boson $\pi$ and the heavy-field are found to depend on a mixture of low- and high-energy modes. Crucially, the gap between these two energies increases as the strength of the turn increases, making high-energy modes more difficult to access at energy scales comparable to the horizon inverse length-scale. As a consequence, although the Goldstone boson stays coupled to the heavy-field, low- and high-energy modes decouple and evolve independently. The end result is a system where only low-energy modes play a relevant role for the generation of curvature perturbations.

\end{itemize} 
Given these characteristics, one may deduce a single-field EFT governing the dynamics at low energy modes (valid at horizon crossing) by integrating out the heavy-field under question.\footnote{For alternative approaches on effective field theories deduced by integrating heavy fields, please see refs.~\cite{Cremonini:2010sv, Jackson:2010cw,  Cremonini:2010ua, Jackson:2011qg, Shiu:2011qw, Avgoustidis:2012yc, Jackson:2012qp, Battefeld:2012qx, Chen:2012ge}.} This turns out to be equivalent to truncate the high-energy modes everywhere in the theory, implying that the heavy-field takes the role of a Lagrange multiplier, to be solved in terms of the Goldstone boson field. The result is a low energy EFT for the Goldstone boson alone, with nontrivial self-interactions leading to interesting properties that differ significantly from those predicted by canonical single-field inflation. For example, a first outstanding property is that the speed of sound $c_s$ at which Goldstone boson perturbations propagate is reduced whenever there is a turn $\dot \theta \neq 0$, with a value determined by the relation
\be
\frac{1}{c_s^{2}} = 1 + \frac{4 \dot \theta^2}{M_{\rm eff}^2} ,
\ee
where $\dot \theta$ and $M_{\rm eff}$ are the quantities already introduced. As shown in~\cite{Cespedes:2012hu}, such an effective field theory remains valid as long as
\be
| \ddot \theta |  \ll M_{\rm eff}  | \dot \theta | ,
\ee
which is a necessary condition ensuring that heavy-fields will not become excited during a turn\footnote{This condition is in fact equivalent to ask the familiar adiabaticity condition $| \dot \omega_{+}  / \omega_+^2 | \ll 1$, where $\omega_{+}$ is the frequency of the high-energy modes implied by the heavy-fields~\cite{Achucarro:2012yr}.}. Furthermore, and consistent with the non-linear realization of the Goldstone boson self-interactions, at small speeds of sounds $c_s^2 \ll 1$ the effective field theory contain sizable cubic self-interactions that inevitably lead to large non-Gaussianity. For instance, at long wavelengths, one find that the EFT is of the form~(\ref{eft-basic}), with $A$ given by
\be
A = -\frac{1}{2}(1 - c_s^2).
\ee
On the other hand, the interaction with a heavy-field may imply the appearance of a {\it  new physics} regime, a range of energy for which the Goldstone boson dispersion relation becomes dominated by a quadratic dependence on the momentum $\omega \sim p^2$~\cite{Baumann:2011su, Achucarro:2012yr, Gwyn:2012mw}. As such, if horizon crossing happened during this regime, the prediction of observables are drastically affected by the {\it new physics} scale dependent operators. This class of EFT's remains weakly coupled all the way up to the cutoff scale at which heavy-fields are allowed to be integrated out~\cite{Baumann:2011su, Gwyn:2012mw}. 

The previous set of findings has paved the way for a more refined understanding of how low energy effective field theories of inflation relate to the ultraviolet parent theories from which they decent. However, there is still much to be learned about the way heavy-fields affect the low energy evolution of adiabatic curvature perturbations. For instance, one may ask how would this picture change if not only one, but several massive fields interacted with the Goldstone boson parametrizing inflation.\footnote{ Fundamental theories such as supergravity and string theory typically predict a large number of scalar fields, most of them expected to remain stabilized (heavy) during inflation. However, since in these theories scalar fields have a geometrical origin, it is still an open challenge to construct models of inflation where all the fields (other than the inflaton) remain stabilized~\cite{Copeland:1994vg, Covi:2008ea, Covi:2008cn, Covi:2008zu, Hardeman:2010fh, Borghese:2012yu}.}  The purpose of this article is to extend the previous body of work by deducing and analyzing the class of single field EFT's obtained in those cases where the Goldstone boson interacted with multiple heavy-fields, all of them representing fluctuations orthogonal to the trajectory. We have two main reasons to pursue this goal: First, we wish to know if the effects of heavy-fields on the low energy dynamics of curvature perturbations increase as the number of heavy-fields increases. In second place, we would like to understand in which way the new couplings, due to additional heavy-fields, would affect the Goldstone boson self-interactions. 

With these two previous motivations in mind, we extend the analysis of a Goldstone boson interacting with a single heavy-field to the case in which it interacts with two heavy-fields. We compute the effective field theory obtained by integrating out the two heavy-fields and analyze the conditions for this limit to remain a fair description of the low energy dynamics of the system. We show that the existence of a third heavy-field indeed may imply larger effects on the low energy dynamics, and that its presence generally induces new self-interactions for the Goldstone boson that are not accounted for in the simpler case of a single heavy-field. Similar to the single-heavy-field case, these new couplings appear whenever the background trajectory in multi-field target space becomes non-geodesic. We find that low energy observables, such as the power spectrum and bispectrum, are sensitive to these couplings, and therefore future observations can be used to discern the number of heavy-fields with which the Goldstone boson interacted during inflation. In particular, we deduce that at long wavelengths, the effective action describing this class of models is of the form~(\ref{eft-basic}), with $A$ generically constrained to be:
\be
A \leqslant -\frac{1}{2}(1 - c_s^2).
\ee
This result implies that, under the assumption that during horizon crossing modes are parametrized by (\ref{eft-basic}), future observations might rule out the existence of interactions between curvature perturbations and a large number of heavy fields.\footnote{Another possibility is that modes crossed the horizon during the new physics regime, in which the Goldstone boson is described in terms of a modified dispersion relation $\omega \sim p^2$. In such case, one is forced to parametrize the period of horizon crossing with a different EFT incorporating operators with nontrivial scalings~\cite{Baumann:2011su, Gwyn:2012mw}.}

We have organized this article as follows: In Section~\ref{Setup} we present the basic setup to be studied and introduce the notation that will be used throughout our work to handle inflationary trajectories traversing a landscape of heavy-fields. In Section~\ref{EFT-derivation}, we analyze the specific case in which the fields orthogonal to the inflationary trajectory are heavy enough that they can be integrated out. We analyze the full multi-field dynamics of this regime and deduce the effective field theory governing the low energy dynamics of the Goldstone boson fluctuations. Then, in Section~\ref{Discussion} we discuss our results by analyzing the observational consequences of the resulting effective field theory for the Goldstone boson. Finally, in Section~\ref{Conclussions} we provide our concluding remarks.

\section{Inflation in a heavy-field landscape} \label{Setup}
\setcounter{equation}{0}

We commence by presenting the basic inflationary setup to be analyzed in the rest of this work. We are interested in studying inflationary systems with three scalar fields $\phi^a (t,x)$ (with $a=1,2,3$) described by a generic action of the form
\be
S_{\rm tot} = \frac{M_{\rm Pl}^2}{2} \int d^4x \sqrt{-g} R + S_{\rm scalar},
\label{action-with-grav}
\ee
where $M_{\rm Pl}$ stands for the Planck mass, $R$ is the Ricci scalar constructed out of the metric $g_{\mu \nu}$ with a $(-,+,+,+)$ signature, and $S_{\rm scalar}$ represents the action for the scalar sector of the theory, given by
\be
S_{\rm scalar} =  - \frac{1}{2}\int d^4x \sqrt{-g} \left[ g^{\mu\nu}\partial_{\mu}\phi^a\partial_{\nu}\phi_a + 2V(\phi)\right] ,
\label{action-scalar}
\ee
where $V (\phi)$ is the scalar field potential. Given that we are interested in a general model-independent analysis, we will not specify the dependance of the potential $V(\phi)$ on the scalar fields $\phi^a$. Instead, we shall only specify local properties of the potential along the background trajectory, consistent with the existence of heavy-fields interacting with the inflaton.

\subsection{Background dynamics}

\begin{figure}[!ht]
\begin{center}
\includegraphics[scale=0.53]{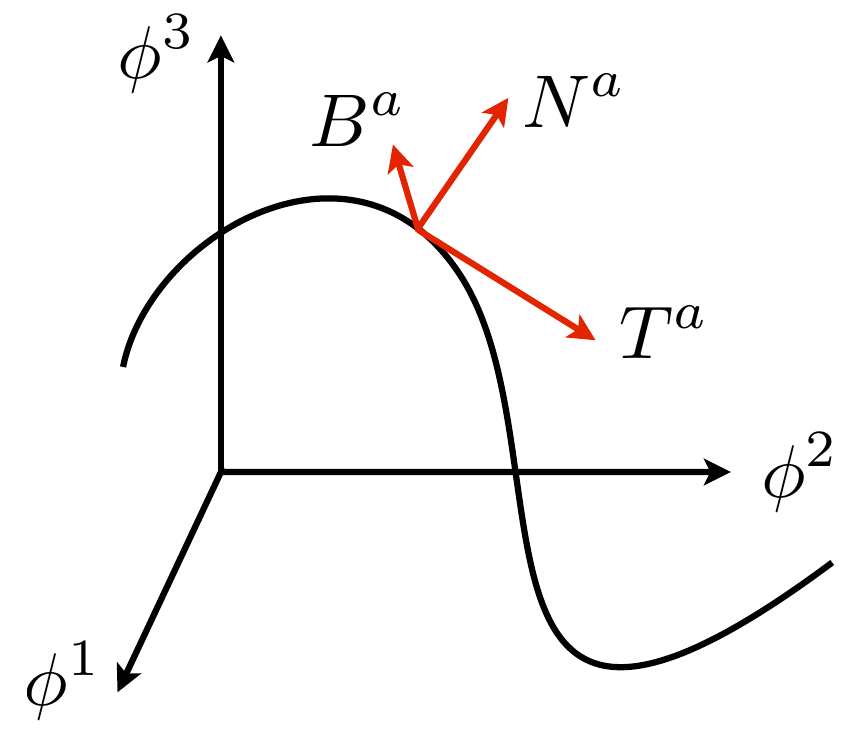}
\caption{A schematic plot of the triad of vectors $\{ T^a, N^a , B^a \}$ defined with respect to the background trajectory $\phi^a_0(t)$. \label{plot-vectors}}
\end{center}
\end{figure}

We assume that the potential $V$ is such that there exist homogeneous time-dependent solutions of the system in which the universe inflates.  This, in turn, means that there exists a background scalar field trajectory in the 3-field target space, parametrized by $t$, hereby denoted by $\phi_0^{a} (t)$. Then, assuming a flat Friedmann-Robertson-Walker background metric of the form $ds^2 = - dt^2 + a^2 d {\bf x}^2$, the background equations of motion determining the trajectory $\phi_0^{a} (t)$ for the scalar fields are given by
\be
\ddot \phi_0^a + 3 H \dot \phi_0^a + V^a = 0, \qquad a = 1,2,3, \label{background-eqs}
\ee
where $H \equiv \dot a/a$ is the usual Hubble expansion rate. These three equations need to be supplemented with Friedmann's equation which, in the present context, is found to be given by
\be
3 H^3 = \frac{1}{M_{\rm Pl}^2} \left(  \frac{1}{2} \dot \phi_0^2 + V  \right), \label{Friedmann-eq}
\ee
where $\dot \phi_0^2 \equiv \delta_{a b} \dot \phi_0^a \dot \phi_0^a$. Putting these two equations together, one deduces an additional equation relating the change of the expansion rate with the rapidity $\dot \phi_0$ of the scalar field along the trajectory:
\be
\dot H = - \frac{\dot \phi_0^2}{2M_{\rm Pl}^2}.
\ee
To study the nontrivial aspects implied by a given path traversing the landscape, it is convenient to define a triad of unit vectors moving along with the trajectory, parametrized by $t$. We choose to work with a standard basis consisting of a tangent vector $T^a$, a normal vector $N^a$ and a binormal vector $B^a$, all of them defined as
\bea
T^a&=&  \dph_0^a / \dph_0 , \label{triad-T} \\
N^a&\propto & \dot T^a ,  \label{triad-N} \\
B^a &\propto& (\delta^a_{b} - T^a T_b ) \dot N^{b} ,  \label{triad-B}
\eea
with positive proportionality coefficients, such that vectors are normalised as $N^a N_a = B^a B_a = T^a T_a = 1$ (we rise and lower indices with $\delta^{ab}$ and $\delta_{ab}$ respectively). These vectors remain mutually orthogonal, and their time evolution may be parametrized by two angular velocities $\dot \theta$ and $\dot \varphi$, defined as:
\bea
\dot{T}^a&=&-\dth N^a ,  \\
\dot N^a&=&\dot \theta  T^a-\dvp B^a  , \\
\dot B^a&=&\dvp  N^a .
\eea
It may be seen that $\dot \theta$ is the rate of change of $T^{a}$ along the direction $- N^{a}$, whereas $\dot \varphi$ is the rate of change of $B^{a}$ along the direction $+N^{a}$. In other words, $\dot \theta$ is the angular velocity of the turning trajectory, whereas $\dot \varphi$ parametrizes how this turn spirals (see Figure \ref{plot-vectors}).
Having introduced this set of vectors~\cite{Gordon:2000hv, GrootNibbelink:2000vx, GrootNibbelink:2001qt}, the background equations of motion (\ref{background-eqs}) may be rewritten by projecting them along the three available directions. One obtains
\bea
\ddot \phi_0 + 3 H \dot \phi_0 + V_T = 0, \label{EOM-proyected-T} \\
\dot \theta = \frac{V_N}{\dot \phi_0} , \\
B^a V_a = 0, 
\eea
where we have defined the projections $V_T \equiv T^a V_a$ and $V_N \equiv N^a V_a$. The first equation (\ref{EOM-proyected-T}) is nothing but the usual equation of motion for a single-field background, with the inflaton rolling down a potential of slope $V_{T}$. Using (\ref{EOM-proyected-T}) and (\ref{Friedmann-eq}) we may now characterise the inflationary dynamics in terms of slow roll parameters as usual. That is, by defining the following dimensionless slow roll parameters
\be
\epsilon = - \frac{\dot H}{H^2} ,  \qquad \eta =  - \frac{\ddot \phi_0}{H \dot \phi_0} , \qquad \xi = - \frac{\dddot \phi_0}{H \ddot \phi_0}, \label{sr-parameters-def}
\ee
one deduces from~(\ref{Friedmann-eq}) and~(\ref{EOM-proyected-T}) the following relations among these quantities,
\bea
\epsilon = \frac{M_{\rm Pl}^2}{2} \left( \frac{V_T}{V} \right)^2 \left( \frac{3 - \epsilon}{3 - \eta} \right)^2 , \label{rewrite-eq-motion-1} \\
3 (\epsilon + \eta) = M_{\rm Pl}^2  \frac{V_{TT}}{V} (3 - \epsilon) + \xi \eta , \label{rewrite-eq-motion-2}
\eea
where $V_{TT} \equiv T^a \nabla_a V_T \equiv T^a \nabla_a ( T^a \nabla_a V)$.
Slow roll inflation will persist as long as $\epsilon \ll 1$, $\eta \ll 1$ and $\xi \ll 1$ hold. With (\ref{rewrite-eq-motion-1}) and (\ref{rewrite-eq-motion-2}) these slow-roll conditions are seen to be equivalent to\footnote{Notice that with definition~(\ref{sr-parameters-def}), the usual $\eta_V$-parameter defined in terms of the second derivative of the potential is given by $\eta_V = \epsilon + \eta$.}
\be
\epsilon = \frac{M_{\rm Pl}^2}{2} \left( \frac{V_T}{V} \right)^2 , \qquad \epsilon + \eta = M_{\rm Pl}^2  \frac{V_{TT}}{V} ,
\ee
which further translates into restrictions on the shape of the potential along the trajectory. At this point, it is very important to emphasise that these slow roll conditions only imply restrictions on background quantities along the trajectory, but tell us nothing about the turns of the trajectory. As discussed in full detail in refs.~\cite{Cespedes:2012hu} and \cite{Achucarro:2012yr}, in the case of two-field models of inflation, it is perfectly possible to have sudden turns with $\dot \theta \gg H$ without implying a violation of the aforementioned  slow-roll conditions. The same arguments can be used to state that, in the case of three-field models of inflation, one can have $\dot \theta \gg H$ and $\dot \varphi \gg H$ simultaneously, without necessarily violating slow-roll whatsoever.

\subsection{Perturbation dynamics}

We now move on to consider perturbations about an arbitrary inflationary trajectory.
A convenient way of studying scalar fluctuations without specifying the inflationary model, is by introducing the Goldstone boson $\pi$ as the fluctuation along the direction of broken time translation symmetry~\cite{Achucarro:2012sm}. In the present context, this is precisely equivalent to define the Goldstone boson as the fluctuation along the inflationary trajectory.\footnote{This is simply because the inflationary trajectory consists of a path parametrized by $t$.} In addition to the Goldstone boson, there are two other scalar field fluctuations, hereby called $\mathcal F_1$ and $\mathcal F_2$, which denote fluctuations away from the trajectory, along the two available directions $N^a$ and $B^a$. The definition of these three scalar fluctuations may be summarized by writing the complete set of scalar fields $\phi^a (t,x)$ in terms of the background fields $\phi^a_0(t)$, and the vectors $N^a(t)$ and $B^a(t)$ as:
\be
\phi^a(t,x) \equiv \phi^a_0(t+\pi)+N^a(t+\pi)\calF_1+B^a(t+\pi)\calF_2 . \label{perturbed-fields}
\ee
Notice that $\pi (t,x)$ appears through the replacement $t \to t + \pi (t,x)$ in the argument of background quantities.  To deal with the gravitational sector, we may adopt the Arnowitt-Deser-Misner (ADM) formalism~\cite{Arnowitt:1962hi} to parametrize space-time, requiring that we write the metric as
\be
ds^2=-N^2dt^2+\gamma_{ij}(N^idt+dx^i)(N^jdt+dx^j) , \label{ADM-metric}
\ee
where $N$ and $N^i$ are the lapse and shift functions (here playing the role of Lagrange multipliers) and $\gamma_{ij}$ is the induced metric describing the 3-D spatial foliations parametrized by $t$. In terms of these quantities, the components of the metric $g_{\mu \nu}$ and its inverse $g^{\mu \nu}$ are given by
\begin{equation}
\begin{split}
g_{00}&=-N^2+\gamma_{ij}N^iN^j,\quad g_{0i}=\gamma_{ij}N^j,\quad g_{ij} = \gamma_{ij}, \\
g^{00}&=-\dfrac{1}{N^2},\quad g^{0i}=\dfrac{N^i}{N^2}\quad g^{ij}=\gamma^{ij}-\dfrac{N^iN^j}{N^2},
\end{split} \label{ADM-parametrization}
\end{equation}
where $\gamma^{ij}$ is the inverse of $\gamma_{ij}$. Moreover, we adopt the flat gauge, in which the spatial metric $\gamma_{ij}$ takes the form:
\be
\gamma_{ij} = a^2 \delta_{ij}.
\ee
To obtain the action for the perturbations, we may now introduce the parametrization (\ref{perturbed-fields}) for $\phi^a(t,x)$ back into the action (\ref{action-with-grav}). The result is given by the following full action, including background fields and fluctuations:
\bea
\label{action-Goldstone-full-pre}
S & = & \int d^4x \frac{Na^3}{2} \bigg\{ -\frac{6 M_{\rm Pl}^2H^2}{N^2} + \frac{4 M_{\rm Pl}^2 H}{N^2}N^i{}_{,i} + \frac{ M_{\rm Pl}^2}{2N^2} \left( N^i{}_{,j}N^j{}_{,i} + \delta_{ij}N^{i,k}N^j{}_{,k} - 2N^i{}_{,i}N^j{}_{,j} \right) 
\nonumber\\
&& + \frac{1}{N^2} \left[ ( \dot\phi_0+ \dot \theta \mathcal F_1 )^2 + \dot \varphi^2 ( {\mathcal F}_1^2 + {\mathcal F}_2^2 )  \right] \left[ \left( 1 + \dot\pi - N^i\pi_{,i} \right)^2 - \frac{N^2}{a^2}\left(\nabla\pi\right)^2 \right]
\nonumber\\
&& + \frac{2 \dot \varphi }{N^2} \left( 1 + \dot\pi - N^i\pi_{,i} \right) \left[  \mathcal F_2 (\dot\calF_1 - N^i\calF_{1 ,i}) -  \mathcal F_1 (\dot\calF_2 - N^i\calF_{2 ,i})  \right] 
\nonumber\\
&& - \frac{2 \dot \varphi }{a^2} \nabla \pi \left[  \mathcal F_2 \nabla \calF_1  -  \mathcal F_1 \nabla \calF_2   \right]   + \frac{1}{N^2} \left( \dot\calF_1 - N^i\calF_{1 ,i} \right)^2 + \frac{1}{N^2} \left( \dot\calF_2 - N^i\calF_{2 ,i} \right)^2
\nonumber\\
&&     - \frac{\left(\nabla\calF_1 \right)^2}{a^2} - \frac{\left(\nabla\calF_2 \right)^2}{a^2}   - 2V\left(\phi_0^a+N^a\calF_1  +B^a\calF_2   \right) \bigg\} \, . 
\eea
To deal with this action, we need to solve the constraint equations for $N$ and $N^i$. To simplify this, we set ourselves to obtain the action for $\pi$, $\mathcal F_1$ and $\mathcal F_2$ only up to cubic order in the fields. This implies that it is only necessary to solve the constraint equations up to linear order in $N-1$ and $N^i$. Then, by writing $N = 1 + \delta N$ and $N^i = \partial^i \psi + v^i$, with $\partial_i v^i = 0$, we find the solutions
\bea
v^i &=& 0 , \\ 
\delta N &=& \epsilon H \pi , \label{delta-N-solution} \\
\frac{\Delta}{a^2} \psi &=& - \epsilon H (\dot \pi - \epsilon H \pi) - \frac{\dot \theta \dot \phi_0}{H M_{\rm Pl}^2 } \mathcal F_1 . \label{N-i-solution}
\eea
Replacing these expressions back into (\ref{action-Goldstone-full-pre}) we obtain the full action for the fluctuations up to cubic order. However, because we are interested in studying inflation in the slow roll retime, where $\epsilon \ll 1$, we are allowed to consider the decoupling limit, where the gravitational effects implied by $\delta N$ and $N^i$ on the evolution of the Goldstone boson become negligible. More specifically, in the regime where the Goldstone boson fluctuations carry energies $\omega \gg \Lambda_{\rm dec} \sim \epsilon H$ one may drop the couplings coming from the constraint solutions (\ref{delta-N-solution}) and (\ref{N-i-solution}), which otherwise imply terms of order $\epsilon$. This step leads us to consider the following action valid at the decoupling limit $\omega \gg \Lambda_{\rm dec}$
\bea
\label{action-Goldstone-full}
S_{\rm dec} & = & \frac{1}{2}  \int d^4x a^3 \bigg\{    ( \dot\phi_0+ \dot \theta \mathcal F_1 )^2 \left[  \dot\pi^2 - \frac{1}{a^2}\left(\nabla\pi\right)^2 \right] +  2 \dot \theta \left( 2 \dot \phi_0 + \dot{\theta} \mathcal F_1 \right) \mathcal F_1  \dot \pi 
\nonumber\\
&& +  \dot \varphi^2 ( {\mathcal F}_1^2 + {\mathcal F}_2^2 )   \left[ 2  \dot\pi +   \dot\pi ^2 - \frac{1}{a^2}\left(\nabla\pi\right)^2 \right]
  + 2 \dot \varphi  \left( 1 + \dot\pi  \right) \left[  \mathcal F_2 \dot\calF_1  -  \mathcal F_1 \dot\calF_2   \right] 
\nonumber\\
&& - \frac{2 \dot \varphi }{a^2} \nabla \pi \left[  \mathcal F_2 \nabla \calF_1  -  \mathcal F_1 \nabla \calF_2   \right]   +   \dot\calF_1^2 +  \dot\calF_2 ^2   - \frac{\left(\nabla\calF_1 \right)^2}{a^2} - \frac{\left(\nabla\calF_2 \right)^2}{a^2} \nn\\ 
&& - \sum_{ij} \mathcal M_{i j } \mathcal F_i \mathcal F_j - \sum_{ij} \mathcal C_{i j k} \mathcal F_i \mathcal F_j \mathcal F_k   \bigg\} \,  ,
\eea
where the mass matrix $\mathcal M^2_{ij}$ is found to have elements given by:
\bea
\mathcal M^2 = \left(\begin{array}{cc} V_{NN} - \dot \theta^2 - \dot \varphi^2  & V_{NB}  \\ V_{NB} & V_{BB}  - \dot \varphi^2   \end{array}\right) . \label{Mass-matrix}
\eea
In this expression, $V_{NB} \equiv N^a B^b V_{a b} \equiv B^a N^b V_{a b}$.  In addition, the cubic term proportional to $\mathcal C_{i j k}$ appears from third derivatives of the potential $V$ away from the inflationary trajectory. 
It is worth noting that at quadratic order the Goldstone boson only interacts with the isocurvature field $\mathcal F_1$, which is precisely due to the parametrisation of the inflationary trajectory in terms of the triad (\ref{triad-T})-(\ref{triad-B}). Because this triad is aligned with respect to the trajectory (and not with respect to the mass matrix of the fields $\mathcal M^2$) in general we expect the existence of non-vanishing off-diagonal terms $\mathcal M_{12}^2 = \mathcal M_{21}^2 \neq 0$.

For completeness, we write down the equations of motion for the fluctuations deduced by varying the action (\ref{action-Goldstone-full}) with  respect to the three fields, $\pi$, $\mathcal F_1$ and $\mathcal F_2$. First, the equation of motion for the Goldstone boson $\pi$ is found to be:
\bea
&& \frac{1}{a^3} \frac{d}{dt}  \left[a^3  \left( ( \dot \phi_0 + \dot \theta \mathcal F_1)^2 + \dot \varphi^2 (\mathcal F_1^2 + \mathcal F_2^2)  \right)  \dot \pi  \right] - \frac{1}{a^2} \nabla  \left[ \left( ( \dot \phi_0 + \dot \theta \mathcal F_1)^2 + \dot \varphi^2 (\mathcal F_1^2 + \mathcal F_2^2)  \right)  \nabla \pi  \right]   \nn  \\ 
&& = - \frac{d}{dt} \left[ \dot \theta \left( 2 \dot \phi_0 + \dot \theta \mathcal F_1 \right) \mathcal F_1 + \dot \varphi^2  (\mathcal F_1^2 + \mathcal F_2^2) + \dot \varphi \left( \mathcal F_2 \dot {\mathcal F_1} - \mathcal F_1 \dot {\mathcal F_2} \right)  \right]  \nn \\  && - \frac{\dot \varphi}{a^2} \nabla \left[ \mathcal F_2 \nabla {\mathcal F_1} - \mathcal F_1 \nabla {\mathcal F_2}  \right]  . \label{EOM-pi}
\eea
The equation of motion for the heavy-field $\mathcal F_1$ is found to be:
\bea
&& \ddot \calF_1 + 3 H \calF_1  - \frac{\nabla^2}{a^2} \calF_1 + \mathcal M_{11}^2 \calF_1 - ( \dot \varphi^2 + \dot \theta^2 ) \left[ 2  \dot\pi +   \dot\pi ^2 - \frac{1}{a^2}\left(\nabla\pi\right)^2 \right] \mathcal F_1   \nn \\&&  
   =   \dot \theta  \dot\phi_0  \left[ 2 \dot \pi +   \dot\pi^2 - \frac{1}{a^2}\left(\nabla\pi\right)^2 \right]   - \mathcal M_{12}^2 \calF_2    - 2 \dot \varphi (1 + \dot \pi) \dot{ \mathcal F}_2  \nn \\&& - 3 H \dot \varphi   \mathcal F_2  -  \ddot \varphi (1 + \dot \pi)  \mathcal F_2 + 2 \frac{ \dot \varphi }{a^2}  \nabla \pi \nabla { \mathcal F}_2  - \dot \varphi \left[  \ddot \pi + 3 H \dot \pi  - \frac{1}{a^2} \nabla^2 \pi  \right]  { \mathcal F}_2 .  \label{EOM-F1}
\eea
And finally, the equation of motion for the heavy-field $\mathcal F_2$ is found to be:
\bea   
&& \ddot \calF_2 + 3 H \calF_2  - \frac{\nabla^2}{a^2} \calF_2 + \mathcal M_{22}^2 \calF_2 - \dot \varphi^2  \left[ 2  \dot\pi +   \dot\pi ^2 - \frac{1}{a^2}\left(\nabla\pi\right)^2 \right] \mathcal F_2  =   - \mathcal M_{21}^2 \calF_1  + 2 \dot \varphi (1 + \dot \pi) \dot{ \mathcal F}_1  \nn \\&&  + 3 H \dot \varphi  \mathcal F_1  +  \ddot \varphi (1 + \dot \pi)  \mathcal F_1 - 2 \frac{ \dot \varphi }{a^2}  \nabla \pi \nabla { \mathcal F}_1 +\dot \varphi \left[ \ddot \pi + 3 H \dot \pi  -  \frac{ 1 }{a^2} \nabla^2 \pi  \right]  { \mathcal F}_1 . \label{EOM-F2}
\eea
In agreement with the analysis of ref.~\cite{Achucarro:2012sm}, the previous equations are consistent with the particular solution $\pi = $constant, and $\mathcal F_1 = \mathcal F_2 = 0$, which is reached shortly after horizon crossing.

\subsection{The linear regime}

We now examine the evolution of fluctuations in the linear regime, paying special attention to their dynamics on sub-horizon scales ({\it i.e.} when the the wavelength of perturbations is shorter than the de Sitter radius $H^{-1}$). Keeping linear terms in eqs. (\ref{EOM-pi})-(\ref{EOM-F2}), and expressing them in Fourier space, we obtain
\bea
  \ddot \pi  + 3 H   \dot \pi  + \frac{k^2}{a^2}  \pi   &=& - \frac{2}{\dot \phi_0}  \left[\dot \theta  \dot { \mathcal F_1 }  +  \ddot \theta  \mathcal F_1  \right] , \\
 \ddot \calF_1 + 3 H \calF_1  + \frac{k^2}{a^2} \calF_1 + \mathcal M_{11}^2 \calF_1 
  & =&   2 \dot \theta  \dot\phi_0  \dot \pi   - \mathcal M_{12}^2 \calF_2    - 2 \dot \varphi  \dot{ \mathcal F}_2   - 3 H \dot \varphi   \mathcal F_2  -  \ddot \varphi \mathcal F_2 , \\
 \ddot \calF_2 + 3 H \calF_2   +\frac{k^2}{a^2} \calF_2 + \mathcal M_{22}^2 \calF_2  &=&   - \mathcal M_{21}^2 \calF_1  + 2 \dot \varphi \dot{ \mathcal F}_1  + 3 H \dot \varphi  \mathcal F_1  +  \ddot \varphi  \mathcal F_1 ,
\eea
where we have also dropped terms suppressed by the slow roll parameters, to stay consistent with the decoupling limit. Recall that the triad $\{ T^a , N^a , B^a \}$ has been chosen so that it remains aligned with the inflationary trajectory, as in eqs.~(\ref{triad-T})-(\ref{triad-B}). As a consequence, at linear order the Goldstone boson $\pi$  remains coupled only to the isocurvature field $\mathcal F_1$, with the strength of the coupling determined by the value of $\dot \theta$. On the other hand, the coupling between the isocurvature mode $\mathcal F_1$ and the binormal mode $\mathcal F_2$ is determined by the combination $\mathcal M_{12}^2 \pm 2  \dot \varphi \partial_t$ (with the sign depending on the field $\partial_t$ acts upon). The mass matrix (\ref{Mass-matrix}) is fixed by the choice of this basis, and any attempt to diagonalize it will change this interaction structure by coupling $\mathcal F_2$ with $\pi$. Thus, in general, we expect a non-vanishing value of $\mathcal M_{12}^2$ even in the absence of spiraling turns ($\dot \varphi = 0$).

To learn more about the kinematical structure of the system, we disregard time derivatives of $\dot \theta$ and $\dot \varphi$ and focus our attention on sub-horizon modes, with $p \equiv k/a \gg H$. Then, the previous equations simplify to
\bea
&&  \ddot \pi_c + p^2 \pi_c   =  - 2   \dot \theta  \dot {\mathcal F} _1 , \label{linear-sub-horizon-1} \\
&& \ddot \calF_1  + p^2 \calF_1 + \mathcal M_{11}^2 \calF_1 
   =   2 \dot \theta   \dot \pi_c   - \mathcal M_{12}^2 \calF_2    - 2 \dot \varphi  \dot{ \mathcal F}_2    , \label{linear-sub-horizon-2} \\
&& \ddot \calF_2  + p^2 \calF_2 + \mathcal M_{22}^2 \calF_2  =   - \mathcal M_{21}^2 \calF_1  + 2 \dot \varphi \dot{ \mathcal F}_1    ,  \label{linear-sub-horizon-3}
\eea
where $\pi_c = \dot \phi_0 \pi$ is the canonically normalised Goldstone boson.
Notice that since $p \equiv k/a \gg H$, one has $| \dot p | / p^2 \ll 1$, implying that we may consider the adiabatic approximation whereby $p$ is treated as a constant. Then, by assuming the ansatz $\pi , \mathcal F_1, \mathcal F_2 \propto e^{- i \omega}$, the previous eqs.~(\ref{linear-sub-horizon-1})-(\ref{linear-sub-horizon-3}) take the form
\be
\Omega \left(\begin{array}{c}  \pi_c  \\ \mathcal F_1 \\  \mathcal F_2 \end{array}\right) = 0 . \label{linear-sub-horizon-array}
\ee
where the frequency matrix $\Omega$ is given by:
\be
\Omega \equiv \left(\begin{array}{ccc} -\omega^2 + p^2 & - 2 i \dot \theta  \omega & 0 \\   2 i \dot \theta  \omega  & -\omega^2 + p^2 + \mathcal M_{11}^2 &  \mathcal M_{12}^2 - 2 i \dot \varphi \omega  \\0 &  \mathcal M_{21}^2 + 2 i \dot \varphi \omega & -\omega^2 + p^2 + \mathcal M_{22}^2  \end{array}\right) . \label{Omega-def}
\ee
To solve these equations, we must demand $\det \Omega = 0$, which determines the following cubic algebraic equation for $\omega$:
\be
(p^2 - \omega^2) (\mathcal M_{12}^4 + 4 \dot \varphi^2 \omega^2) - (\mathcal M_{22}^2 + p^2 - \omega^2) \left( p^2 \mathcal M_{11}^2 + p^4 - (\mathcal M_{11}^2 + 2 p^2 + 4 \dot \theta^2 ) \omega^2 + \omega^4 \right) = 0. \label{charact-eq}
\ee
Even though in this section we are interested in studying the system at sub-horizon scales, it is instructive to analyze the equation (\ref{charact-eq}) by its own merits, and explore the limit $p\rightarrow 0$ (as if the system were embedded in a Minkowski background).
For $p = 0$ one of the solutions corresponds to the case $\omega = 0$. This is consistent with the fact that $\pi = \,\,$constant and $\mathcal F_1 = \mathcal F_2 = 0$ is a solution of the system, and implies that there is a massless mode (to be identified as the Goldstone boson mode). Then, expressing the three frequencies about $p = 0$, we find\footnote{Here it should be understood that, even though we are expanding the solutions of (\ref{charact-eq}) about $p=0$, these are strictly valid as long as $ H^2 \ll \omega^2$.}
\bea
\omega_{\rm light}^2 &=&c_s^2  p^2 + \frac{\left(1 - c_s^2 \right)^2 }{ \det \mathcal M^2 c_s^{-2}  }   \left[   \mathcal M_{22}^2 + \frac{ \mathcal M_{12}^4}{ \mathcal M_{22}^2}  -  \frac{4 c_s^2 \dot \varphi^2}{1 - c_s^2}  \right] p^4  + \mathcal O (p^6)  , \label{omega-0} \\
\omega_{\rm I}^2 &=& \frac{ {\rm tr}  \mathcal M^2 + 4 ( \dot \theta^2 + \dot \varphi^2 )}{2} - \frac{1}{2}  \sqrt{ \left[ {\rm tr}  \mathcal M^2 + 4 ( \dot \theta^2 + \dot \varphi^2 ) \right]^2  - 4 \, { \rm det} \mathcal M^2 \, c_s^{-2} }    , \qquad  \label{omega-1} \\
\omega_{\rm II}^2 &=& \frac{ {\rm tr}  \mathcal M^2 + 4 ( \dot \theta^2 + \dot \varphi^2 )}{2} + \frac{1}{2}  \sqrt{ \left[ {\rm tr}  \mathcal M^2 + 4 ( \dot \theta^2 + \dot \varphi^2 ) \right]^2  - 4 \, { \rm det} \mathcal M^2 \, c_s^{-2}}   , \qquad  \label{omega-2} 
\eea
where we have defined the speed of sound $c_s$ via the relation:
\bea
\frac{1}{c_s^2} =1+\frac{4 \dth^2 \mathcal M_{22}^2}{\det \mathcal M^2 } .
\eea
Notice that we have dropped the $p$-dependence of $\omega_{\rm I}$ and $\omega_{\rm II}$ for simplicity.\footnote{These $p$-dependent contributions are in fact suppressed in the low energy regime where $\omega_{\rm light}^2 \ll \omega_{\rm I, II}^2$ to be studied in the next section.} In addition, notice that $\omega_{\rm I} \leqslant \omega_{\rm II}$ by definition. A direct check of these relations shows that in the limit $\mathcal M_{12}^2 = \dot \varphi^2 = 0$ we recover the case in which only one massive field interacts with the Goldstone boson~\cite{Achucarro:2012yr}:
\bea
\omega_{\rm light}^2 &=& c_s^2  p^2 + \left(1 - c_s^2 \right)^2 \frac{  p^4 }{ \mathcal M_{11}^2 c_s^{-2} } + \mathcal O (p^6)  \label{omegas}   \\
\omega_{\rm I}^2 &=& \mathcal M_{11}^2 + 4 \dot \theta^2 = \mathcal M_{11}^2 c_s^{-2}\\
\omega_{\rm II}^2 &=& \mathcal M_{22}^2 ,
\eea
where we have assumed $\mathcal M_{11}^2 + 4 \dot \theta^2  < \mathcal M_{22}^2$ for definiteness (otherwise we would have obtained the inverted relations $\omega_{\rm I}^2 = \mathcal M_{22}^2$ and $\omega_{\rm II}^2 = \mathcal M_{11}^2 + 4 \dot \theta^2$).

In general, we see that the coupled system of equations (\ref{linear-sub-horizon-array}) imply that the fields $\pi$, $\mathcal F_1$ and $\mathcal F_2$ are linear combinations of modes with frequencies $\omega_{\rm light}$, $\omega_{\rm I}$ and $\omega_{\rm II}$ in the following form
\bea
\pi &=& \pi_{\rm light} e^{ - i \omega_{\rm light} t } + \pi_{\rm I} e^{ - i \omega_{\rm I} t } + \pi_{\rm II} e^{ - i \omega_{\rm II} t } , \\
\mathcal F_1  &=& \mathcal F_{\textrm{ 1-light}} e^{ - i \omega_{\rm light} t } + \mathcal F_{\rm 1I} e^{ - i \omega_{\rm I} t } + \mathcal F_{\rm 1II} e^{ - i \omega_{\rm II} t } , \\
\mathcal F_2  &=& \mathcal F_{\textrm{ 2-light}} e^{ - i \omega_{\rm light} t } + \mathcal F_{\rm 2I} e^{ - i \omega_{\rm I} t } + \mathcal F_{\rm 2II} e^{ - i \omega_{\rm II} t } .
\eea
The amplitudes $\pi_{\rm light}$, $\pi_{\rm I}$, $\pi_{\rm II}$, $\mathcal F_{\textrm{ 1-light}} $, $\mathcal F_{\rm 1I} $, $\mathcal F_{\rm 1II}$, $\mathcal F_{\textrm{ 2-light}}$, $ \mathcal F_{\rm 2I}$ and $\mathcal F_{\rm 2II} $ are all functions of $p$, and determined trivially by (\ref{linear-sub-horizon-array}) except three normalization coefficients, that may be fixed by quantizing the theory. In the particular case where the inflationary trajectory is not subject to turns ({\it i.e.} $\dot \theta = \dot \varphi = 0$),  the matrix of eq.~(\ref{linear-sub-horizon-array}) becomes diagonal, and only $\pi_{\rm light}$, $\mathcal F_{\rm 1I} $, and $\mathcal F_{\rm 2II} $ remain non-vanishing. In such a case, assuming that $\mathcal M_{11}^2 \leqslant \mathcal M_{22}^2$, the frequencies reduce to 
\be
\omega_{\rm light}^2 = p^2 , \qquad \omega_{\rm I}^2 =  \mathcal M_{11}^2 , \qquad \omega_{\rm II}^2 =  \mathcal M_{22}^2,
\ee
and there is a one to one correspondence between frequencies and fields. However, it is important to emphasize that in the presence of turns ({\it i.e.} $\dot \theta \neq 0 $ and $\dot \varphi \neq 0$) there will always be a mixing between fields and modes, implying non-trivial consequences for the dynamics of the low energy Goldstone boson, as we shall verify in the following section.

\section{Effective field theory} \label{EFT-derivation}
\setcounter{equation}{0}

In the previous section we analysed the dynamics of inflationary systems with three scalar fields, which may be understood in terms of a Goldstone boson interacting with two massive scalar fields. We now move on to consider the case in which these two massive fields remain heavy, and therefore contribute with heavy degrees of freedom to the particle content of the theory. Such a regime exists only for wavelengths such that the frequency of the light mode is found to be much smaller than the frequencies of the two heavy degrees of freedom:
\be
\omega_{\rm light}^2 \ll \omega_{\rm I}^2 \leqslant \omega_{\rm II}^2. \label{frequency-hierarchy}
\ee
As long as this condition is satisfied, the creation of high-energy quanta of energies $\omega_{\rm I}$ and $\omega_{\rm II}$ will remain kinematically precluded to processes involving low-energy degrees of freedom characterized by $\omega_{\rm light}$. Thus $\omega_{\rm I}$ constitutes the cut-off energy scale defining the validity of the  effective field theory for low energy modes of frequency $\omega_{\rm light}$. However, because $\omega_{\rm I}$ and  $\omega_{\rm II}$ depend on time-dependent background quantities, eq.~(\ref{frequency-hierarchy}) needs to be complemented with the additional adiabaticity conditions~\cite{Achucarro:2012yr} 
\be
\frac{| \dot \omega_{\rm I} |}{\omega_{\rm I}^2} \ll 1, \qquad \frac{| \dot \omega_{\rm II} |}{\omega_{\rm II}^2} \ll 1 ,
\ee
ensuring that high-frequency quanta will not be excited by strong sudden turns of the background inflationary trajectory. 

\subsection{Preliminaries}

In what follows we deduce the effective field theory describing the dynamics of the low energy modes characterized by the frequency $\omega_{\rm light}$, subject to the hierarchy (\ref{frequency-hierarchy}). First, because there is a large number of parameters involved in the definition of both $\omega_{\rm I}^2$ and $\omega_{\rm II}^2$, we need to make some simplifying assumptions about their values. To start with, we assume that both $\omega_{\rm I}^2$ and $\omega_{\rm II}^2$ are of the same order. By inspecting eqs.~(\ref{omega-1}) and (\ref{omega-2}) we see that this condition implies that the cutoff scale is of order
\bea
 \left[ {\rm tr}  \mathcal M^2 + 4 ( \dot \theta^2 + \dot \varphi^2 ) \right]^2  \sim  4 \, { \rm det} \mathcal M^2 \, c_s^{-2} .
\eea
in order to avoid a hierarchy between $\omega_{\rm I}^2$ and $\omega_{\rm II}^2$. In second place, we only consider inflationary trajectories where  $\dot \varphi$ is at most of order $\dot \theta$.\footnote{Notice that although both $\dot\theta$ and $\dot\varphi$ have a geometrical interpretation, in principle there are no constraints on how large the ratio $|\dot\varphi/\dot\theta|$ can be.} Putting together these two assumptions, one finds that both $\mathcal M_{22}^2$ and $\mathcal M_{11}^2  + 4 \dot \theta^2 $ are of the same order as the cutoff scale $\Lambda_{\rm UV}^2 = \omega_{\rm I}^2$ of the effective field theory:
\be
  \mathcal M_{22}^2   \sim \mathcal M_{11}^2  + 4 \dot \theta^2 \sim \Lambda_{\rm UV}^2 .\label{Lambda-parameters}
\ee
It is important to realize that under the present assumptions, $\mathcal M_{11}^2$ and $\dot \theta^2$ are not necessarily of the same order, and a hierarchy among their values is perfectly possible~\cite{Cespedes:2012hu}.

Next, we may anticipate the range of validity of the low energy EFT in terms of the momentum carried by the fluctuations. For this, we see that the EFT will remain valid as long as $\omega_{\rm light}^2 \ll \Lambda_{\rm UV}^2$. Then, noticing from~(\ref{omegas})  and~(\ref{Lambda-parameters}) that the dispersion relation for the light mode is of the general form
\be
\omega_{\rm light}^2 \sim c_s^2  p^2 +  \frac{ \left(1 - c_s^2 \right)^2 }{\Lambda_{\rm UV}^2 } p^4  + \mathcal O (p^6 / \Lambda_{\rm UV}^4) ,  \label{light-generic}
\ee
we see that, independently of the value of $c_s$, the effective field theory is valid as long as the wavelength 
\be
p^2 \ll \Lambda_{\rm UV}^2 .  \label{p-UV}
\ee

Finally, we argue that the term proportional to $\dot \varphi^2 $ appearing in the light mode dispersion relation~(\ref{omega-0}), is always subleading when compared to any other term in the expression. Indeed, from eq.~(\ref{light-generic}) we see that the contribution quartic in $p$ dominates only if $c_s^2 \ll 1$, in which case the contribution due to  $\dot \varphi^2 $ will be suppressed by a factor $c_s^2$ against the remaining term $ \mathcal M_{22}^2 +  \mathcal M_{12}^4 /  \mathcal M_{22}^2 $ (recall that we are taking $\dot \varphi^2$ at most of order $\sim \dot \theta^2$). Thus, we are allowed to take 
\be
\omega_{\rm light}^2  = c_s^2  p^2 + \frac{\left(1 - c_s^2 \right)^2 }{ \det \mathcal M^2 c_s^{-2}  }   \left[   \mathcal M_{22}^2 + \frac{ \mathcal M_{12}^4}{ \mathcal M_{22}^2}   \right] p^4  + \mathcal O (p^6 / \Lambda_{\rm UV}^4 )  , \label{disp-rel-1}
\ee
as the dispersion relation for the light mode, with terms of order $\mathcal O (p^6) $ always subleading~\cite{Gwyn:2012mw}.

\subsection{Computation of the effective field theory}

We are now ready to compute the desired effective field theory. We will do this by expressing both heavy-fields, $\mathcal F_1$ and $\mathcal F_2$ in terms of the light Goldstone boson $\pi$, with the help of the equations of motion (\ref{EOM-F1}) and (\ref{EOM-F2}). The following two considerations will help in this task:
\begin{itemize}

\item We first notice that the absence of $\dot \varphi$ in the dispersion relation (\ref{disp-rel-1}) allows us to drop any term containing $\dot \varphi$ in (\ref{EOM-F1}) and (\ref{EOM-F2}), as long as it is linear in the fields. However, we must keep $\dot \varphi$ in those terms which are of higher order in the fields. 

\item In addition, the modified dispersion relation (\ref{disp-rel-1}) is consistent  with $\omega^2 \ll p^2 + \mathcal M_{11}^2$ and $\omega^2 \ll p^2 + \mathcal M_{22}^2$ for all values of $p$ up to the cutoff scale $\Lambda_{\rm UV}$. This means that we can drop the second time derivatives $\ddot {\mathcal F_{1}}$ and $\ddot {\mathcal F_{2}}$ in the equations of motion  (\ref{EOM-F1}) and (\ref{EOM-F2}) respectively. 

\end{itemize}
To appreciate the relevance of these two points more clearly, we may analyze their effects when applied to the linear equations of motion (\ref{linear-sub-horizon-array}) valid at sub-horizon scales. In this case, the matrix $\Omega$ is found to be:
\be
\Omega \equiv \left(\begin{array}{ccc} - \omega^2 +  p^2 & - 2 i \dot \theta  \omega & 0 \\   2 i \dot \theta  \omega  &  p^2 + \mathcal M_{11}^2 &  \mathcal M_{12}^2 \\0 &  \mathcal M_{21}^2  & p^2 + \mathcal M_{22}^2  \end{array}\right) , \label{Omega-approx}
\ee
from where it is straightforward to deduce the following  dispersion relation for the light mode:
\bea
\omega_{\rm light}^2 &=&c_s^2  p^2 + \frac{\left(1 - c_s^2 \right) }{ \det \mathcal M^2 c_s^{-2}  }   \left[   \mathcal M_{22}^2 + \frac{ \mathcal M_{12}^4}{ \mathcal M_{22}^2}  \right] p^4 + \mathcal O (p^6)  . \label{omega-0-EFT}
\eea
The only difference between this expression and that found in (\ref{disp-rel-1}) is a missing extra factor $(1 - c_s^2)$ in front of the quartic term of (\ref{omega-0-EFT}). This comes from having neglected the second time derivatives of the heavy-fields (see ref.~\cite{Gwyn:2012mw} for a detailed explanation of this in the case of a single heavy-field). However, this difference is marginal, as the quartic term is only relevant if the speed of sound is suppressed ($c_s^2 \ll 1$).

Next, we write the equations of motion at most linear in the heavy-fields $\mathcal F_1$ and $\mathcal F_2$, but to quadratic order in $\pi$ (this will allow us to consistently deduce an EFT action for $\pi$ valid to cubic order in $\pi$):
\bea
&&   \left[ - \frac{\nabla^2}{a^2} + \mathcal M_{11}^2  - 2  \dot\pi  ( \dot \varphi^2 + \dot \theta^2 ) \right]  \calF_1    
   =   \dot \theta  \dot\phi_0  \left[ 2 \dot \pi +   \dot\pi^2 - \frac{1}{a^2}\left(\nabla\pi\right)^2 \right]  - \mathcal M_{12}^2 \calF_2      , \\
&& \left[ - \frac{\nabla^2}{a^2}  + \mathcal M_{22}^2  - 2  \dot\pi \dot \varphi^2  \right]  \calF_2 =   - \mathcal M_{21}^2 \calF_1   .
\eea
Since we are neglecting second order time derivatives, these equations may be interpreted as constraint equations for the Lagrange multipliers $\mathcal F_1$ and $\mathcal F_2$. As such, they automatically provide the low energy evolution of the heavy-fields $\mathcal F_1$ and $\mathcal F_2$ as sourced by the Goldstone boson $\pi$. The solution to these equations are given by
\bea
&&    \calF_1
   =   \Omega_2 \frac{ \dot \theta  \dot\phi_0}{ \Omega_1\Omega_2 -  \mathcal M_{21}^4 }  \left[ 2 \dot \pi +   \dot\pi^2 - \frac{1}{a^2}\left(\nabla\pi\right)^2 \right]  , \label{sol-F1} \\
&& \calF_2 =  -  \dot \theta  \dot\phi_0  \frac{ \mathcal M_{21}^2 }{\Omega_1 \Omega_2 -  \mathcal M_{21}^4 }\left[ 2 \dot \pi +   \dot\pi^2 - \frac{1}{a^2}\left(\nabla\pi\right)^2 \right] , \label{sol-F2}
\eea
where the operators $\Omega_1$ and $\Omega_2$ are defined as:
\bea
\Omega_1 &\equiv &  \left[ - \frac{\nabla^2}{a^2} + \mathcal M_{11}^2  - 2  \dot\pi  ( \dot \varphi^2 + \dot \theta^2 ) \right]  , \\
\Omega_2 &\equiv &  \left[ - \frac{\nabla^2}{a^2}  + \mathcal M_{22}^2  - 2  \dot\pi \dot \varphi^2  \right] .
\eea
Replacing the solutions (\ref{sol-F1}) and (\ref{sol-F2}) back into the full action (\ref{action-Goldstone-full}), and consistently dropping those terms in the action the led to disregarded terms in the equations of motion, we are led to the single-field Goldstone-boson action in the decoupling limit:
\bea
\label{action-Goldstone-full-2}
S_{\rm EFT} & = & \frac{1}{2}  \int d^4x a^3 \dot\phi_0^2 \bigg\{      \left[  \dot\pi^2 - \frac{1}{a^2}\left(\nabla\pi\right)^2 \right]  + 4  \dot \theta^2   \dot \pi     \frac{\mathcal M_{22}^2 - \nabla^2/a^2  }{  (\mathcal M_{11}^2 - \nabla^2/a^2 ) (\mathcal M_{22}^2 - \nabla^2/a^2 ) -  \mathcal M_{21}^4 } \dot \pi    \nn\\ 
&& + 8 \dot \theta^2 \dot \varphi^2  \left[ \dot \pi  \frac{\mathcal M_{12}^2  }{  (\mathcal M_{11}^2 - \nabla^2/a^2 ) (\mathcal M_{22}^2 - \nabla^2/a^2 ) -  \mathcal M_{21}^4 } \right]^2 \dot \pi   \nn\\ 
&& + 8 \dot \theta^2 ( \dot \theta^2 + \dot \varphi^2 ) \left[ \dot \pi  \frac{(\mathcal M_{22}^2 - \nabla^2/a^2 )  }{  (\mathcal M_{11}^2 - \nabla^2/a^2 ) (\mathcal M_{22}^2 - \nabla^2/a^2 ) -  \mathcal M_{21}^4 } \right]^2 \dot \pi    \nn\\ 
&& + 2  \dot \theta^2    \left[   \dot\pi^2 - \frac{1}{a^2}\left(\nabla\pi\right)^2 \right]    \frac{\mathcal M_{22}^2 - \nabla^2/a^2  }{  (\mathcal M_{11}^2 - \nabla^2/a^2 ) (\mathcal M_{22}^2 - \nabla^2/a^2 ) -  \mathcal M_{21}^4 }  \dot \pi    \nn\\ 
&& +  2 \dot \theta^2  \dot \pi   \frac{ \mathcal M_{22}^2 - \nabla^2/a^2   }{ (\mathcal M_{11}^2 - \nabla^2/a^2 )(\mathcal M_{22}^2 - \nabla^2/a^2 ) -  \mathcal M_{21}^4 }  \left[    \dot\pi^2 - \frac{1}{a^2}\left(\nabla\pi\right)^2 \right]   \bigg\} \, .
\eea
This action may be further simplified by recalling that our formalism only allows us to integrate heavy fields at   wavelengths such that (\ref{p-UV}) is respected. This allows us to write:
\bea
\label{action-Goldstone-full-extra}
S_{\rm EFT} & = & \frac{1}{2}  \int d^4x a^3 \dot\phi_0^2 \bigg\{      \left[  \dot\pi^2 - \frac{1}{a^2}\left(\nabla\pi\right)^2 \right]  +   \dot \pi     \frac{4  \dot \theta^2  }{  \det \mathcal M^2 / \mathcal M_{22}^2- \nabla^2/a^2    } \dot \pi    \nn\\ 
&& + \frac{1}{2}  \left( 1 + \frac{\dot \varphi^2}{\dot \theta^2}  \frac{ \mathcal M_{22}^4 + \mathcal M_{12}^4  }{ \mathcal M_{22}^4 } \right) \left[ \dot \pi  \frac{ 4 \dot \theta^2 }{ \det \mathcal M^2 / \mathcal M_{22}^2 -  \nabla^2/a^2  } \right]^2 \dot \pi    \nn\\ 
&& + \frac{1}{2}    \left[   \dot\pi^2 - \frac{1}{a^2}\left(\nabla\pi\right)^2 \right]    \frac{4  \dot \theta^2  }{ \det \mathcal M^2 / \mathcal M_{22}^2 -  \nabla^2/a^2  }   \dot \pi    \nn\\ 
&& +  \frac{1}{2} \dot \pi  \frac{ 4 \dot \theta^2 }{ \det \mathcal M^2 / \mathcal M_{22}^2 -  \nabla^2/a^2  }   \left[    \dot\pi^2 - \frac{1}{a^2}\left(\nabla\pi\right)^2 \right]   \bigg\} \, .
\eea
This action constitutes one of our main results. It summarises the effect of two heavy-fields on the evolution of a single adiabatic mode, parametrized by the Goldstone boson mode $\pi$. The dispersion relation for the Goldstone boson mode may be read from the quadratic part of the action, and is found to be given by:
\be
\omega^2 = \frac{ (\mathcal M_{11}^2 + p^2 ) \mathcal M_{22}^2  -  \mathcal M_{21}^4}{(\mathcal M_{11}^2 + p^2 ) \mathcal M_{22}^2 -  \mathcal M_{21}^4 +  4  \dot \theta^2 \mathcal M_{22}^2 } p^2 .
\ee
Expanding this expression in powers of $p^2$, we obtain back the dispersion relation (\ref{omega-0-EFT}). It may be seen that at energies larger than
\be
\Lambda_{\rm new}^2 \sim \Lambda_{\rm UV}^2 c_s ,
\ee
the dispersion relation changes from a linear dependence on the momentum $\omega \propto p$ to a quadratic dependence $\omega \propto p^2$. This regime has been dubbed new physics regime~\cite{Baumann:2011su}, and it signals the regime where the non-trivial contributions due to the Laplacian $\nabla^2$ become important in (\ref{action-Goldstone-full-2}).

\section{Discussion} \label{Discussion}
\setcounter{equation}{0}

We now wish to highlight and discuss some of the main characteristics emerging from the effective field theory deduced in the previous section. First of all, the form of action (\ref{action-Goldstone-full-extra}) coincides with that studied in ref.~\cite{Gwyn:2012mw}, where general arguments about the effects of heavy fields on curvature perturbations where given. There, the non trivial effects coming from heavy physics was parametrized by a single mass scale $M$, representing the mass of a single heavy field modifying the kinematics of the low energy Goldstone boson. Direct comparison between both approaches allows us to identify $M$ in terms of the entries of the mass matrix $\mathcal M^2$ as:
\be
M^2 = \det \mathcal M^2 / \mathcal M_{22}^2.
\ee
In addition, motivated by the EFT parametrization of ref.~\cite{Cheung:2007st}, in ref.~\cite{Gwyn:2012mw} the Goldstone boson self couplings were parametrized with the help of a set of the couplings $M_n^4$, where $n$ denoted the order of expansion of the EFT in terms of the Newtonian potential $g^{00} + 1$. In the present case, it is direct to read that the relation between $M_3^4$ and $M_2^4$ is given by
\bea
\frac{M_3^4}{M_2^4}=-\frac{3}{4}(c_s^{-2}-1)   \left(  1   + \frac{ \dot \varphi^2 }{ \dot \theta^2}  \frac{\mathcal M_{22}^4 + \mathcal M_{12}^4 }{\mathcal M_{22}^4 }    \right) , \label{M3-M2}
\eea
where one sees that all the nontrivial effects due to the presence of the second field are due to the ratio $\dot \varphi^2 / \dot \theta^2$.

There are two relevant limits of the deduced effective field theory, depending on the value of $\Lambda_{\rm new}^2 = \Lambda_{\rm UV}^2 c_s$ relative to $H^2$. If $H^2 \ll \Lambda_{\rm new}^2 $, the dispersion relation takes the simple form $\omega = c_s p$ during horizon crossing, and the relevant EFT action parametrizing this process becomes:
\bea
\label{action-Goldstone-full-3}
S_{\rm EFT} & = & \frac{1}{2}  \int d^4x a^3 \dot\phi_0^2 \bigg\{      \left[  \frac{1}{c_s^2} \dot\pi^2 - \frac{1}{a^2}\left(\nabla\pi\right)^2 \right]   +      \left(  \frac{1}{c_s^2} - 1 \right) \dot \pi  \left[    \dot\pi^2 - \frac{1}{a^2}\left(\nabla\pi\right)^2 \right]    \nn\\ 
&& + \frac{  1  }{2   }  \left(  \frac{1}{c_s^2} - 1 \right)^2 \left(   1   + \frac{ \dot \varphi^2 }{ \dot \theta^2}  \frac{\mathcal M_{22}^4 + \mathcal M_{12}^4 }{\mathcal M_{22}^4 }    \right)  \dot \pi^3  \bigg\} \,  .
\eea
This form of the action may be compared with the one found in the original EFT analysis of ref.~\cite{Cheung:2007st}. In terms of the parametrization offered by eq.~(\ref{eft-basic}) one deduces that:
\be
A = -\frac{1}{2}(1 - c_s^2)   \left(  1   + \frac{ \dot \varphi^2 }{ \dot \theta^2}  \frac{\mathcal M_{22}^4 + \mathcal M_{12}^4 }{\mathcal M_{22}^4 }    \right) .
\ee
Thus we see that a second heavy-field enlarges the EFT parameter encountered in the single field case. Crucially, this happens only in one direction, and we are able to conclude that the generic effect implied by a second field is to allow the inequality:
\be
A \leqslant -\frac{1}{2}(1 - c_s^2).
\ee
This inequality may be tested, by means of the relations~\cite{Senatore:2009gt}:
\bea
f^{\mathrm{eq}}_{\mathrm{NL}}  &=&  \frac{1-c_s^2}{c_s^2} (- 0.276 + 0.0785 A) , \label{eq-long}  \\
f^{\mathrm{orth}}_{\mathrm{NL}}  &=& \frac{1-c_s^2}{c_s^2} ( 0.0157 - 0.0163 A ) , \label{orth-long}
\eea 
as long as large non-Gaussian signatures are observed.

Next, we may consider the limit in which horizon crossing happens during the new physics regime. Here  the dispersion relation is dominated by a quadratic term ($\omega \sim p^2$), and one is forced to consider the full action (\ref{action-Goldstone-full-extra}). This form of the action was studied in detail in ref.~\cite{Baumann:2011su, Gwyn:2012mw} where it was noticed that the nontrivial scale dependence implied by the insertions $4 \dot \theta^2 / (M^2 - \nabla^2 / a^2)$ (with $M^2 = \det \mathcal M^2 / \mathcal M_{22}^2$), would modify drastically the computation of observables in terms of background inflationary quantities. For instance, the quadratic part of the action (\ref{action-Goldstone-full-extra}) in the new physics regime reads 
\be
\label{action-Goldstone-full-new-ph}
S_{\rm EFT}^{(2)}  =  \frac{1}{2}  \int d^4x a^3 \dot\phi_0^2 \bigg\{      \left[  \dot\pi^2 - \frac{1}{a^2}\left(\nabla\pi\right)^2 \right]  -   \dot \pi     \frac{4  \dot \theta^2  }{  \nabla^2/a^2    } \dot \pi \bigg\} ,
\ee
from where one deduces that the power spectrum $\mathcal P_{\zeta}$, and the tensor to scalar ratio $r$, are given by:
\be
\mathcal P_{\zeta}  \simeq \frac{5.4}{100} \frac{H^2}{M_{\rm Pl}^2 \epsilon} \sqrt{\frac{\dot \theta}{H}} , \qquad   r \simeq 3.8 \epsilon  \sqrt{\frac{H}{\dot \theta} } .
\ee
A detailed characterization of the shape of non-Gaussianity in the new physics regime is still missing, but it is possible to infer that the size of equilateral non-Gaussianity is of order
\be
 f_{\rm NL} \sim \frac{\dot \theta}{H},
\ee
with its actual value determined by $M_3^4$ given by (\ref{M3-M2}).

\section{Conclussions} \label{Conclussions}
\setcounter{equation}{0}

We have deduced the effective field theory describing the evolution of curvature perturbations during inflation, in the specific case where the Goldstone boson mode interacted with two heavy fields. Our main result is summarized by eq.~(\ref{action-Goldstone-full-extra}), which provides the explicit effective field action for the Goldstone boson field $\pi$. Crucially, the couplings induced by the presence of a second heavy-field are distinguishable from those appearing in the single heavy-field case, implying that a detailed characterization of non-Gaussianities will allow us constrain this class of scenarios.  In particular, the presence of a second field implies the following general inequality involving the parameters $c_s$ and $A$, appearing in the EFT of eq.~(\ref{eft-basic})
\be
A \leqslant -\frac{1}{2}(1 - c_s^2), \label{ineq-conclusions}
\ee
which is saturated in the single-field case. In terms of the angular velocities $\dot \theta$ and $\dot \varphi$ parametrizing the multi-field inflationary trajectory, such an inequality becomes stronger as $\dot \varphi$ becomes of the same order than $\dot \theta$ (which may be as large as the cutoff scale $\Lambda_{\rm UV}$).

Our results represent a significant step towards a better understanding of the collective effects that many heavy fields may have on the evolution of adiabatic perturbations, and highlight the importance of using effective field theory techniques to interpret future observations. In particular, our results show, in an eloquent manner, how different values of $c_s$ and $A$ correspond to different, and potentially distinguishable, UV realizations of inflation. Several outstanding questions remain to be answered: For instance, in the present analysis, we have assumed that the parameter space is such that the two high frequencies $\omega_{\rm I}$ and $\omega_{\rm II}$ are of the same order, simplifying the derivation of the desired effective field theory. Thus, it would be desirable to study the system in other limits allowed by the parameters, such as $\omega_{\rm light}^2 \ll \omega_{\rm I}^2 \ll \omega_{\rm II}^2$, and/or $\dot \varphi^2 \gg \dot \theta^2$. Also, given that one generically expects several heavy fields to have interacted with curvature perturbations during inflation, it would be important to know whether the inclusion of additional heavy fields would modify inequality (\ref{ineq-conclusions}).  Last but not least, it would be interesting to study the way in which a second heavy field would generate features in the primordial spectra due to possible sudden turns of the inflationary trajectory in the landscape.

\section*{Acknowledgments}

We are grateful to Ana Ach\'ucarro, Jinn-Ouk Gong, Subodh Patil and Spyros Sypsas for useful comments and discussions. This work was supported by a Fondecyt project number 1130777 and a Conicyt ``Anillo" project (ACT1122). GAP would like to thank the kind hospitality of Cambridge University, CERN and Kings College London while this work was being completed.


\begin{thebibliography}{99}


\bibitem{Guth:1980zm} 
  A.~H.~Guth,
  ``The Inflationary Universe: A Possible Solution to the Horizon and Flatness Problems,''
  Phys.\ Rev.\ D {\bf 23}, 347 (1981).

\bibitem{Linde:1981mu}
 A.~D.~Linde,
 ``A New Inflationary Universe Scenario: A Possible Solution Of The Horizon,
 Flatness, Homogeneity, Isotropy And Primordial Monopole Problems,''
 Phys.\ Lett.\  B {\bf 108} (1982) 389.

\bibitem{Albrecht:1982wi}
 A.~Albrecht and P.~J.~Steinhardt,
 ``Cosmology For Grand Unified Theories With Radiatively Induced Symmetry
 Breaking,''
 Phys.\ Rev.\ Lett.\  {\bf 48} (1982) 1220.



\bibitem{Mukhanov:1981xt}
 V.~F.~Mukhanov and G.~V.~Chibisov,
 ``Quantum Fluctuation And Nonsingular Universe. (In Russian),''
 JETP Lett.\  {\bf 33} (1981) 532
 [Pisma Zh.\ Eksp.\ Teor.\ Fiz.\  {\bf 33} (1981) 549].
 

\bibitem{Komatsu:2010fb} 
  E.~Komatsu {\it et al.}  [WMAP Collaboration],
  ``Seven-Year Wilkinson Microwave Anisotropy Probe (WMAP) Observations: Cosmological Interpretation,''
Astrophys.\ J.\ Suppl.\  {\bf 192}, 18 (2011)
[arXiv:1001.4538 [astro-ph.CO]].

\bibitem{Hlozek:2011pc} 
  R.~Hlozek, J.~Dunkley, G.~Addison, J.~W.~Appel, J.~R.~Bond, C.~S.~Carvalho, S.~Das and M.~Devlin {\it et al.},
  ``The Atacama Cosmology Telescope: a measurement of the primordial power spectrum,''
  Astrophys.\ J.\  {\bf 749}, 90 (2012)
  [arXiv:1105.4887 [astro-ph.CO]].

\bibitem{Bennett:2012fp}
  C.~L.~Bennett, D.~Larson, J.~L.~Weiland, N.~Jarosik, G.~Hinshaw, N.~Odegard, K.~M.~Smith and R.~S.~Hill {\it et al.},
  ``Nine-Year Wilkinson Microwave Anisotropy Probe (WMAP) Observations: Final Maps and Results,''
  arXiv:1212.5225 [astro-ph.CO].





\bibitem{Abell:2009aa} 
  P.~A.~Abell {\it et al.}  [LSST Science and LSST Project Collaborations],
  ``LSST Science Book, Version 2.0,''
  arXiv:0912.0201 [astro-ph.IM].

\bibitem{Schlegel:2011wb} 
  D.~Schlegel {\it et al.}  [BigBOSS Collaboration],
  ``The BigBOSS Experiment,''
  arXiv:1106.1706 [astro-ph.IM].

\bibitem{Amendola:2012ys} 
  L.~Amendola {\it et al.}  [Euclid Theory Working Group Collaboration],
  ``Cosmology and fundamental physics with the Euclid satellite,''
  arXiv:1206.1225 [astro-ph.CO].


\bibitem{21cm}
  M.~McQuinn, O.~Zahn, M.~Zaldarriaga, L.~Hernquist and S.~R.~Furlanetto,
  ``Cosmological parameter estimation using 21 cm radiation from the epoch of reionization,''
  Astrophys.\ J.\  {\bf 653}, 815 (2006)
  [astro-ph/0512263].




\bibitem{Martin:2004yi}
  J.~Martin and C.~Ringeval,
  ``Exploring the superimposed oscillations parameter space,''
  JCAP {\bf 0501}, 007 (2005)
  [hep-ph/0405249]~;
  
\bibitem{Shafieloo:2006hs}  
  A.~Shafieloo, T.~Souradeep, P.~Manimaran, P.~K.~Panigrahi and R.~Rangarajan,
  ``Features in the Primordial Spectrum from WMAP: A Wavelet Analysis,''
  Phys.\ Rev.\ D {\bf 75}, 123502 (2007)
  [astro-ph/0611352]~;
  
\bibitem{Hamann:2007pa}  
  J.~Hamann, L.~Covi, A.~Melchiorri and A.~Slosar,
  ``New Constraints on Oscillations in the Primordial Spectrum of Inflationary Perturbations,''
  Phys.\ Rev.\ D {\bf 76}, 023503 (2007)
  [astro-ph/0701380~]~;
  
\bibitem{Hamann:2009bz}    
  J.~Hamann, A.~Shafieloo and T.~Souradeep,
  ``Features in the primordial power spectrum? A frequentist analysis,''
  JCAP {\bf 1004}, 010 (2010)
  [arXiv:0912.2728 [astro-ph.CO]].


\bibitem{Starobinsky:1992ts} 
  A.~A.~Starobinsky,
  ``Spectrum of adiabatic perturbations in the universe when there are singularities in the inflation potential,''
  JETP Lett.\  {\bf 55}, 489 (1992)
  [Pisma Zh.\ Eksp.\ Teor.\ Fiz.\  {\bf 55}, 477 (1992)].

\bibitem{Polarski:1992dq}
  D.~Polarski and A.~A.~Starobinsky,
  ``Spectra of perturbations produced by double inflation with an intermediate matter dominated stage,''
  Nucl.\ Phys.\  B {\bf 385}, 623 (1992).

\bibitem{Chung:1999ve} 
  D.~J.~H.~Chung, E.~W.~Kolb, A.~Riotto and I.~I.~Tkachev,
  ``Probing Planckian physics: Resonant production of particles during inflation and features in the primordial power spectrum,''
  Phys.\ Rev.\ D {\bf 62}, 043508 (2000)
  [hep-ph/9910437].

\bibitem{Adams:2001vc} 
  J.~A.~Adams, B.~Cresswell and R.~Easther,
  ``Inflationary perturbations from a potential with a step,''
  Phys.\ Rev.\ D {\bf 64}, 123514 (2001)
  [astro-ph/0102236].

\bibitem{Gong:2005jr} 
  J.~-O.~Gong,
  ``Breaking scale invariance from a singular inflaton potential,''
  JCAP {\bf 0507}, 015 (2005)
  [astro-ph/0504383].

\bibitem{Ashoorioon:2006wc}
 A.~Ashoorioon and A.~Krause,
 ``Power Spectrum and Signatures for Cascade Inflation,''
 hep-th/0607001.

\bibitem{Romano:2008rr} 
  A.~E.~Romano and M.~Sasaki,
  ``Effects of particle production during inflation,''
  Phys.\ Rev.\ D {\bf 78}, 103522 (2008)
  [arXiv:0809.5142 [gr-qc]].


\bibitem{Ashoorioon:2008qr}
 A.~Ashoorioon, A.~Krause and K.~Turzynski,
 ``Energy Transfer in Multi Field Inflation and Cosmological Perturbations,''
 JCAP {\bf 0902}, 014 (2009)
 [arXiv:0810.4660 [hep-th]].


\bibitem{Tye:2008ef}
  S.~-H.~H.~Tye, J.~Xu, Y.~Zhang,
  ``Multi-field Inflation with a Random Potential,''
  JCAP {\bf 0904}, 018 (2009).
  [arXiv:0812.1944 [hep-th]].

\bibitem{Tye:2009ff}
  S.~-H.~H.~Tye, J.~Xu,
  ``A Meandering Inflaton,''
  Phys.\ Lett.\  {\bf B683}, 326-330 (2010).
  [arXiv:0910.0849 [hep-th]].

\bibitem{Barnaby:2010ke} 
  N.~Barnaby,
  ``On Features and Nongaussianity from Inflationary Particle Production,''
  Phys.\ Rev.\ D {\bf 82}, 106009 (2010)
  [arXiv:1006.4615 [astro-ph.CO]].

\bibitem{Achucarro:2010da}
  A.~Achucarro, J.~O.~Gong, S.~Hardeman, G.~A.~Palma and S.~P.~Patil,
  ``Features of heavy physics in the CMB power spectrum,''
  JCAP {\bf 1101} (2011) 030
  [arXiv:1010.3693 [hep-ph]].
  
\bibitem{Chen:2011zf}
  X.~Chen,
  ``Primordial Features as Evidence for Inflation,''
  JCAP {\bf 1201} (2012) 038
  [arXiv:1104.1323 [hep-th]].

\bibitem{Park:2012rh} 
  M.~Park and L.~Sorbo,
  ``Sudden variations in the speed of sound during inflation: features in the power spectrum and bispectrum,''
  arXiv:1201.2903 [astro-ph.CO].

\bibitem{Cespedes:2012hu} 
  S.~Cespedes, V.~Atal and G.~A.~Palma,
  ``On the importance of heavy fields during inflation,''
  JCAP {\bf 1205}, 008 (2012)
  [arXiv:1201.4848 [hep-th]].

\bibitem{Chluba:2012we} 
  J.~Chluba, A.~L.~Erickcek and I.~Ben-Dayan,
  ``Probing the inflaton: Small-scale power spectrum constraints from measurements of the CMB energy spectrum,''
  Astrophys.\ J.\  {\bf 758}, 76 (2012)
  [arXiv:1203.2681 [astro-ph.CO]].

\bibitem{Battefeld:2013xka} 
  T.~Battefeld, J.~C.~Niemeyer and D.~Vlaykov,
  arXiv:1302.3877 [astro-ph.CO].

\bibitem{Jackson:2013mka} 
  M.~G.~Jackson, B.~Wandelt and F.~o.~Bouchet,
  ``Angular Correlation Functions for Models with Logarithmic Oscillations,''
  arXiv:1303.3499 [hep-th].















\bibitem{Linde:1996gt} 
  A.~D.~Linde and V.~F.~Mukhanov,
  ``Nongaussian isocurvature perturbations from inflation,''
  Phys.\ Rev.\ D {\bf 56}, 535 (1997)
  [astro-ph/9610219].

\bibitem{Bartolo:2001cw} 
  N.~Bartolo, S.~Matarrese and A.~Riotto,
  ``Nongaussianity from inflation,''
  Phys.\ Rev.\ D {\bf 65}, 103505 (2002)
  [hep-ph/0112261].


\bibitem{Bernardeau:2002jy} 
  F.~Bernardeau and J.~-P.~Uzan,
  ``NonGaussianity in multifield inflation,''
  Phys.\ Rev.\ D {\bf 66}, 103506 (2002)
  [hep-ph/0207295].
  
\bibitem{Maldacena:2002vr} 
  J.~M.~Maldacena,
  ``Non-Gaussian features of primordial fluctuations in single field inflationary models,''
  JHEP {\bf 0305}, 013 (2003)
  [astro-ph/0210603].
  
\bibitem{Lyth:2002my} 
  D.~H.~Lyth, C.~Ungarelli and D.~Wands,
  ``The Primordial density perturbation in the curvaton scenario,''
  Phys.\ Rev.\ D {\bf 67}, 023503 (2003)
  [astro-ph/0208055].

\bibitem{Alishahiha:2004eh}
  M.~Alishahiha, E.~Silverstein and D.~Tong,
  ``DBI in the sky,''
  Phys.\ Rev.\  D {\bf 70}, 123505 (2004)
  [arXiv:hep-th/0404084].


\bibitem{Bartolo:2004if}
  N.~Bartolo, E.~Komatsu, S.~Matarrese {\it et al.},
  ``Non-Gaussianity from inflation: Theory and observations,''
  Phys.\ Rept.\  {\bf 402}, 103-266 (2004).
  [astro-ph/0406398].

\bibitem{Chen:2006nt} 
  X.~Chen, M.~-x.~Huang, S.~Kachru and G.~Shiu,
  ``Observational signatures and non-Gaussianities of general single field inflation,''
  JCAP {\bf 0701}, 002 (2007)
  [hep-th/0605045].

\bibitem{Chen:2009we}
  X.~Chen and Y.~Wang,
  ``Large non-Gaussianities with Intermediate Shapes from Quasi-Single Field Inflation,''
  Phys.\ Rev.\  D {\bf 81} (2010) 063511
  [arXiv:0909.0496 [astro-ph.CO]].









\bibitem{Review}
For recent reviews on cosmic inflation, see for instance: 
  W.~H.~Kinney,
  ``TASI Lectures on Inflation,''
  arXiv:0902.1529 [astro-ph.CO];
  D.~Baumann,
  ``TASI Lectures on Inflation,''
  arXiv:0907.5424 [hep-th];
  D.~Langlois,
  ``Lectures on inflation and cosmological perturbations,''
  Lect.\ Notes Phys.\  {\bf 800}, 1 (2010)
  [arXiv:1001.5259 [astro-ph.CO]].


\bibitem{Cheung:2007st}
  C.~Cheung, P.~Creminelli, A.~L.~Fitzpatrick, J.~Kaplan and L.~Senatore,
  ``The Effective Field Theory of Inflation,''
  JHEP {\bf 0803} (2008) 014
  [arXiv:0709.0293 [hep-th]].

\bibitem{Weinberg:2008hq}
  S.~Weinberg,
  ``Effective Field Theory for Inflation,''
  Phys.\ Rev.\  D {\bf 77} (2008) 123541
  [arXiv:0804.4291 [hep-th]].
  

\bibitem{Burgess:2003zw}
  C.~P.~Burgess, J.~M.~Cline and R.~Holman,
  ``Effective field theories and inflation,''
  JCAP {\bf 0310}, 004 (2003)
  [hep-th/0306079].
  


\bibitem{Cheung:2007sv} 
  C.~Cheung, A.~L.~Fitzpatrick, J.~Kaplan and L.~Senatore,
  ``On the consistency relation of the 3-point function in single field inflation,''
  JCAP {\bf 0802}, 021 (2008)
  [arXiv:0709.0295 [hep-th]].


\bibitem{Senatore:2009gt} 
  L.~Senatore, K.~M.~Smith and M.~Zaldarriaga,
  ``Non-Gaussianities in Single Field Inflation and their Optimal Limits from the WMAP 5-year Data,''
JCAP {\bf 1001}, 028 (2010)
[arXiv:0905.3746 [astro-ph.CO]].






\bibitem{Senatore:2009cf} 
  L.~Senatore and M.~Zaldarriaga,
  ``On Loops in Inflation,''
  JHEP {\bf 1012}, 008 (2010)
  [arXiv:0912.2734 [hep-th]].
  
\bibitem{Senatore:2010jy} 
  L.~Senatore and M.~Zaldarriaga,
  ``A Naturally Large Four-Point Function in Single Field Inflation,''
  JCAP {\bf 1101}, 003 (2011)
  [arXiv:1004.1201 [hep-th]].

\bibitem{Senatore:2010wk} 
  L.~Senatore and M.~Zaldarriaga,
  ``The Effective Field Theory of Multifield Inflation,''
  JHEP {\bf 1204}, 024 (2012)
  [arXiv:1009.2093 [hep-th]].


\bibitem{Creminelli:2010qf} 
  P.~Creminelli, G.~D'Amico, M.~Musso, J.~Norena and E.~Trincherini,
  ``Galilean symmetry in the effective theory of inflation: new shapes of non-Gaussianity,''
  JCAP {\bf 1102}, 006 (2011)
  [arXiv:1011.3004 [hep-th]].

\bibitem{Baumann:2011dt} 
  D.~Baumann, L.~Senatore and M.~Zaldarriaga,
  ``Scale-Invariance and the Strong Coupling Problem,''
  JCAP {\bf 1105}, 004 (2011)
  [arXiv:1101.3320 [hep-th]].

\bibitem{Baumann:2011su}
  D.~Baumann, D.~Green,
  ``Equilateral Non-Gaussianity and New Physics on the Horizon,''
  JCAP {\bf 1109}, 014 (2011).
  [arXiv:1102.5343 [hep-th]].

\bibitem{Baumann:2011nk} 
  D.~Baumann and D.~Green,
  ``Signatures of Supersymmetry from the Early Universe,''
  Phys.\ Rev.\ D {\bf 85}, 103520 (2012)
  [arXiv:1109.0292 [hep-th]].

\bibitem{Baumann:2011nm} 
  D.~Baumann and D.~Green,
  ``Supergravity for Effective Theories,''
  JHEP {\bf 1203}, 001 (2012)
  [arXiv:1109.0293 [hep-th]].

\bibitem{LopezNacir:2011kk} 
  D.~Lopez Nacir, R.~A.~Porto, L.~Senatore and M.~Zaldarriaga,
  ``Dissipative effects in the Effective Field Theory of Inflation,''
  JHEP {\bf 1201}, 075 (2012)
  [arXiv:1109.4192 [hep-th]].

\bibitem{Baumann:2011ws} 
  D.~Baumann and D.~Green,
  ``A Field Range Bound for General Single-Field Inflation,''
  arXiv:1111.3040 [hep-th].

\bibitem{Noumi:2012vr} 
  T.~Noumi, M.~Yamaguchi and D.~Yokoyama,
  ``Effective field theory approach to quasi-single field inflation,''
  arXiv:1211.1624 [hep-th].
  
\bibitem{Achucarro:2012sm} 
  A.~Achucarro, J.~-O.~Gong, S.~Hardeman, G.~A.~Palma and S.~P.~Patil,
  ``Effective theories of single field inflation when heavy fields matter,''
  JHEP {\bf 1205}, 066 (2012)
  [arXiv:1201.6342 [hep-th]].



\bibitem{Senatore:2012wy} 
  L.~Senatore and M.~Zaldarriaga,
  ``A Note on the Consistency Condition of Primordial Fluctuations,''
  arXiv:1203.6884 [astro-ph.CO].

\bibitem{Behbahani:2012be} 
  S.~R.~Behbahani and D.~Green,
  ``Collective Symmetry Breaking and Resonant Non-Gaussianity,''
  JCAP {\bf 1211}, 056 (2012)
  [arXiv:1207.2779 [hep-th]].

\bibitem{Gwyn:2012mw} 
  R.~Gwyn, G.~A.~Palma, M.~Sakellariadou and S.~Sypsas,
  ``Effective field theory of weakly coupled inflationary models,''
  arXiv:1210.3020 [hep-th].

\bibitem{Achucarro:2012fd} 
  A.~Achucarro, J.~-O.~Gong, G.~A.~Palma and S.~P.~Patil,
  ``Correlating features in the primordial spectra,''
  arXiv:1211.5619 [astro-ph.CO].









\bibitem{Tolley:2009fg}
  A.~J.~Tolley and M.~Wyman,
  ``The Gelaton Scenario: Equilateral non-Gaussianity from multi-field dynamics,''
  Phys.\ Rev.\  D {\bf 81} (2010) 043502
  [arXiv:0910.1853 [hep-th]].

\bibitem{Achucarro:2010jv} 
  A.~Achucarro, J.~-O.~Gong, S.~Hardeman, G.~A.~Palma and S.~P.~Patil,
  ``Mass hierarchies and non-decoupling in multi-scalar field dynamics,''
  Phys.\ Rev.\ D {\bf 84}, 043502 (2011)
  [arXiv:1005.3848 [hep-th]].

\bibitem{Gao:2012uq}
  X.~Gao, D.~Langlois and S.~Mizuno,
  ``Influence of heavy modes on perturbations in multiple field inflation,''
  arXiv:1205.5275 [hep-th].

\bibitem{Pi:2012gf} 
  S.~Pi and M.~Sasaki,
  ``Curvature Perturbation Spectrum in Two-field Inflation with a Turning Trajectory,''
  JCAP {\bf 1210}, 051 (2012)
  [arXiv:1205.0161 [hep-th]].

\bibitem{Achucarro:2012yr} 
  A.~Achucarro, V.~Atal, S.~Cespedes, J.~-O.~Gong, G.~A.~Palma and S.~P.~Patil,
  ``Heavy fields, reduced speeds of sound and decoupling during inflation,''
  Phys.\ Rev.\  D {\bf 86}, 121301(R) (2012)
  [arXiv:1205.0710 [hep-th]].


\bibitem{Saito:2012pd} 
  R.~Saito, M.~Nakashima, Y.~-i.~Takamizu and J.~'i.~Yokoyama,
  ``Resonant Signatures of Heavy Scalar Fields in the Cosmic Microwave Background,''
  JCAP {\bf 1211}, 036 (2012)
  [arXiv:1206.2164 [astro-ph.CO]].







\bibitem{Copeland:1994vg} 
  E.~J.~Copeland, A.~R.~Liddle, D.~H.~Lyth, E.~D.~Stewart and D.~Wands,
  ``False vacuum inflation with Einstein gravity,''
  Phys.\ Rev.\ D {\bf 49}, 6410 (1994)
  [astro-ph/9401011].

\bibitem{Covi:2008ea} 
  L.~Covi, M.~Gomez-Reino, C.~Gross, J.~Louis, G.~A.~Palma and C.~A.~Scrucca,
  ``de Sitter vacua in no-scale supergravities and Calabi-Yau string models,''
  JHEP {\bf 0806}, 057 (2008)
  [arXiv:0804.1073 [hep-th]].

\bibitem{Covi:2008cn} 
  L.~Covi, M.~Gomez-Reino, C.~Gross, J.~Louis, G.~A.~Palma and C.~A.~Scrucca,
  ``Constraints on modular inflation in supergravity and string theory,''
  JHEP {\bf 0808}, 055 (2008)
  [arXiv:0805.3290 [hep-th]].
  
\bibitem{Covi:2008zu} 
  L.~Covi, M.~Gomez-Reino, C.~Gross, G.~A.~Palma and C.~A.~Scrucca,
  ``Constructing de Sitter vacua in no-scale string models without uplifting,''
  JHEP {\bf 0903}, 146 (2009)
  [arXiv:0812.3864 [hep-th]].

\bibitem{Hardeman:2010fh} 
  S.~Hardeman, J.~M.~Oberreuter, G.~A.~Palma, K.~Schalm and T.~van der Aalst,
  ``The everpresent eta-problem: knowledge of all hidden sectors required,''
  JHEP {\bf 1104}, 009 (2011)
  [arXiv:1012.5966 [hep-ph]].
  
\bibitem{Borghese:2012yu} 
  A.~Borghese, D.~Roest and I.~Zavala,
  ``A Geometric bound on F-term inflation,''
  JHEP {\bf 1209}, 021 (2012)
  [arXiv:1203.2909 [hep-th]].


  




\bibitem{Cremonini:2010sv}
  S.~Cremonini, Z.~Lalak and K.~Turzynski,
  ``On Non-Canonical Kinetic Terms and the Tilt of the Power Spectrum,''
  Phys.\ Rev.\  D {\bf 82}, 047301 (2010)
  [arXiv:1005.4347 [hep-th]]~;

\bibitem{Jackson:2010cw} 
  M.~G.~Jackson and K.~Schalm,
  ``Model Independent Signatures of New Physics in the Inflationary Power Spectrum,''
  Phys.\ Rev.\ Lett.\  {\bf 108}, 111301 (2012)
  [arXiv:1007.0185 [hep-th]]~;
  
\bibitem{Cremonini:2010ua} 
  S.~Cremonini, Z.~Lalak and K.~Turzynski,
  ``Strongly Coupled Perturbations in Two-Field Inflationary Models,''
  JCAP {\bf 1103}, 016 (2011)
  [arXiv:1010.3021 [hep-th]]~;
  
\bibitem{Jackson:2011qg} 
  M.~G.~Jackson and K.~Schalm,
  ``Model-Independent Signatures of New Physics in Slow-Roll Inflation,''
  arXiv:1104.0887 [hep-th].
  
\bibitem{Shiu:2011qw} 
  G.~Shiu and J.~Xu,
  ``Effective Field Theory and Decoupling in Multi-field Inflation: An Illustrative Case Study,''
  Phys.\ Rev.\ D {\bf 84}, 103509 (2011)
  [arXiv:1108.0981 [hep-th]]~;
  
\bibitem{Avgoustidis:2012yc} 
  A.~Avgoustidis, S.~Cremonini, A.~C.~Davis, R.~H.~Ribeiro, K.~Turzynski and S.~Watson,
  ``Decoupling Survives Inflation: A Critical Look at Effective Field Theory Violations During Inflation,''
  arXiv:1203.0016 [hep-th].

\bibitem{Jackson:2012qp} 
  M.~G.~Jackson,
  ``Integrating out Heavy Fields in Inflation,''
  arXiv:1203.3895 [hep-th].

\bibitem{Battefeld:2012qx} 
  D.~Battefeld, T.~Battefeld and S.~Schulz,
  JCAP {\bf 1206}, 034 (2012)
  [arXiv:1203.3941 [hep-th]].

\bibitem{Chen:2012ge} 
  X.~Chen and Y.~Wang,
  ``Quasi-Single Field Inflation with Large Mass,''
  arXiv:1205.0160 [hep-th]~;




\bibitem{Gordon:2000hv}
  C.~Gordon, D.~Wands, B.~A.~Bassett and R.~Maartens,
  ``Adiabatic and entropy perturbations from inflation,''
  Phys.\ Rev.\  D {\bf 63} (2000) 023506
  [arXiv:astro-ph/0009131]~;
  
\bibitem{GrootNibbelink:2000vx}
 S.~Groot Nibbelink and B.~J.~W.~van Tent,
 ``Density perturbations arising from multiple field slow-roll inflation,''
 [arXiv:hep-ph/0011325]~;
 
\bibitem{GrootNibbelink:2001qt}
  S.~Groot Nibbelink and B.~J.~W.~van Tent,
  ``Scalar perturbations during multiple field slow-roll inflation,''
  Class.\ Quant.\ Grav.\  {\bf 19} (2002) 613
  [arXiv:hep-ph/0107272].



\bibitem{Arnowitt:1962hi} 
  R.~L.~Arnowitt, S.~Deser and C.~W.~Misner,
  ``The Dynamics of general relativity,''
  gr-qc/0405109.










\end{thebibliography}
\end{document}